\documentclass[aps,pra,reprint,amsmath,amssymb,superscriptaddress,nofootinbib,longbibliography,floatfix]{revtex4-2}
\usepackage{graphicx}
\usepackage[utf8]{inputenc}
\usepackage[T1]{fontenc}
\usepackage{xcolor}
\usepackage{soul}

\usepackage{booktabs}

\usepackage{amssymb}

\usepackage{bm}
\usepackage{mathtools}
\usepackage{chemformula}
\usepackage{hyperref}
\begin{document}

	\title{Crystal structure, magnetic properties and magnetocaloric performance of RE$_{5}$Rh$_2$In$_4$ (RE = Gd-Tm) compounds}

	\author{Altifani Rizky Hayyu}
	\email{altifani.hayyu@doctoral.uj.edu.pl}
	\affiliation{Jagiellonian University, Doctoral School of Exact and Natural Sciences, Faculty of Physics, Astronomy and Applied Computer Science, prof. Stanis\l{}awa \L{}ojasiewicza 11, PL-30-348 Krak\'ow, Poland}
	\affiliation{Jagiellonian University, Faculty of Physics, Astronomy and Applied Computer Science, M.~Smoluchowski Institute of Physics,
		prof. Stanis\l{}awa \L{}ojasiewicza 11, PL-30-348 Krak\'ow, Poland}	
	\author{Stanis\l{}aw Baran}
	\email{stanislaw.baran@uj.edu.pl}
	\affiliation{Jagiellonian University, Faculty of Physics, Astronomy and Applied Computer Science, M.~Smoluchowski Institute of Physics,
		prof. Stanis\l{}awa \L{}ojasiewicza 11, PL-30-348 Krak\'ow, Poland}
	\author{Aleksandra Deptuch}
	\affiliation{Institute of Nuclear Physics, Polish Academy of Sciences, Radzikowskiego 152, PL-31-342 Krak\'ow, Poland}
	\author{Andrzej Szytu\l{}a}
	\affiliation{Jagiellonian University, Faculty of Physics, Astronomy and Applied Computer Science, M.~Smoluchowski Institute of Physics,
		prof. Stanis\l{}awa \L{}ojasiewicza 11, PL-30-348 Krak\'ow, Poland}

\date{\today}
	
\begin{abstract}

Polycrystalline samples of the \ch{RE5Rh2In4} (RE~=~Gd--Tm) intermetallics have been investigated by means of X-ray diffraction (XRD)
as well as by DC and AC magnetometric measurements. The XRD data confirm that the compounds crystallize with the orthorhombic 
Lu$_{5}$Ni$_2$In$_4$-type structure (space group \textit{Pbam}, No.~55). With decreasing temperature, \ch{RE5Rh2In4}
spontaneously order magnetically with the critical temperatures of magnetic order equal to 10.6, 14.2, 15.4, 7.2, 5.9 and 4.9~K
for RE = Gd, Tb, Dy, Ho, Er and Tm, respectively.
The compounds have complex magnetic properties, showing features characteristic of both ferro- and antiferromagnetic orderings.
Moreover, the magnetic properties change with increasing number of the 4f electrons from predominantly ferromagnetic in
Gd$_{5}$Rh$_2$In$_4$ to predominantly antiferromagnetic in Tm$_{5}$Rh$_2$In$_4$. The magnetic data indicate that only the rare
earth atoms carry magnetic moments.
The maximum magnetic entropy change ($-\Delta S_{M}^{max}$) at the 0-9~T magnetic flux density change ($\Delta \mu_0 H$)
equals 12.4~J$\cdot$kg$^{-1}\cdot$K$^{-1}$ at 17~K for Gd$_{5}$Rh$_2$In$_4$, 11.3~J$\cdot$kg$^{-1}\cdot$K$^{-1}$ at 28~K for Tb$_{5}$Rh$_2$In$_4$,
13.1~J$\cdot$kg$^{-1}\cdot$K$^{-1}$ at 19~K for Dy$_{5}$Rh$_2$In$_4$, 16.4~J$\cdot$kg$^{-1}\cdot$K$^{-1}$ at 12~K for Ho$_{5}$Rh$_2$In$_4$,
15.3~J$\cdot$kg$^{-1}\cdot$K$^{-1}$ at 8~K for Er$_{5}$Rh$_2$In$_4$ and 12.6~J$\cdot$kg$^{-1}\cdot$K$^{-1}$ at 6.5~K for Tm$_{5}$Rh$_2$In$_4$. For a selected rare earth element (RE),
the member of the \ch{RE5Rh2In4} family of compounds reaches the highest $-\Delta S_{M}^{max}$ value, when compared
with its RE$_{5}$T$_2$In$_4$ (T = Ni, Pd, Pt) isostructural analogues, making the \ch{RE5Rh2In4} intermetallics
a~good choice for application in low-temperature magnetic refrigeration.

		\bigskip
		\noindent \textbf{keywords}: rare earth intermetallics, magnetic order, magnetic susceptibility, magnetization, magnetocaloric effect, magnetic entropy change
		
	\end{abstract}
	
	\maketitle
	
	\section{Introduction}
	\label{intro}

Rare earth intermetallics have attracted researchers’ interest for many decades because of their intriguing physical properties,
such as complex/frustrated magnetic structures, metamagnetism, large magnetocaloric effect (MCE), valence
fluctuation, heavy fermion behavior, and Kondo effect. Among these compounds, the ones with multiple magnetic sublattices are of
special interest as individual magnetic sublattices may be characterized by individual propagation vectors, individual directions of magnetic
moments and individual critical temperatures of magnetic ordering. Often additional temperature-induced order-order magnetic transitions appear
in this class of materials.

The magnetocaloric performance phenomenon, in which certain materials experience temperature changes in response to an externally applied magnetic field, has attracted wide interest because of the potential in magnetic refrigerant 
applications. Recent investigations on the magnetic properties and magnetocaloric effects of certain rare-earth intermetallics can be found in both review articles and publications focusing on individual compounds~\cite{GUPTA2015562, 
PATINO2023414496, LI2020153810, Guo2021, BHATTACHARYYA20121239, ZHANG2009396, doi:10.1063/1.3130090, ZHANG20112602, PhysRevB.74.132405, doi:10.1063/5.0006281, Zhang_2009, DESOUZA201611, CWIK20181088, BARAN2020106837, 
MatsumotoMagnetocaloric, doi:10.1063/1.2919079, RAJIVGANDHI2018351, NIKITIN1991166, Zou_Jun-Ding_2007, WADA1999689, Burzo2010, Singh_2007, DUC2002873, samantha2006magnetocaloric_effect, C5RA06970J, MAJI2018236, WU2019168, Rawat_2006, 
R_Rawat_2001, SAHU2022103327, Zhang_2015, Mo_2015, HUO20181044, SHEN20112949, xu2012magnetocaloricHoPdAl, Sharma_2018, SHARMA2018317, SHARMA201956, YAO201937, doi:10.1063/1.3253729, s24196326, TALIK2009L30, SHANG2020166055, 
VONRANKE2001970, ZHEN2024176251}.



Among ternary intermetallics, the RE$_{5}$T$_2$In$_4$ family of compounds, where RE -- rare earth element, T -- $n$d transition metal element
(T = Ni~\cite{zaremba1991crystal}, Rh~\cite{Zaremba2008}, Pd~\cite{Sojka200890}, Pt~\cite{zaremba2007}) are of special interest due to their intriguing
magnetic properties. These compounds crystallize in the orthorhombic structure of the Lu$_{5}$Ni$_2$In$_4$-type (\textit{Pbam} space group, No. 55), where rare 
earth atoms occupy three distinct Wyckoff sites: the 2a site and two 4g sites with unique atomic positional parameters~\cite{zaremba1991crystal}.
It is worth noting that the Lu$_{5}$Ni$_2$In$_4$ crystal structure is characteristic of compounds with general composition of RE$_{m+n}$T$_{2n}$X$_{m}$,
where RE -- rare earth element, T -- transition metal element and X -- p-electron element, while $m$ and $n$ value represent the numbers of the
REIn (CsCl-type) and RET$_2$ (AlB$_2$-type) slabs forming the structure.


In the RE$_{5}$Ni$_2$In$_4$ system, Tb$_5$Ni$_2$In$_4$ exhibits ferromagnetic/ferrimagnetic behavior below $T_C=125$~K with an additional
antiferromagnetic component appearing below 20~K~\cite{Ritter_2015}, while Dy$_5$Ni$_2$In$_4$ has ferromagnetic-like ordering below 105~K,
followed by multiple magnetic phase transitions at lower temperatures~\cite{provino2012crystal}. Ho$_5$Ni$_2$In$_4$ orders antiferromagnetically below the Néel temperature of 
$T_N=31$~K and transforms to ferromagnetic/ferrimagnetic structure at $T_C=25$~K~\cite{Ritter_2015}, though another study reports $T_C=19$~K~\cite{zhang2018investigation}.
A different paper reports ferromagnetic ordering in Ho$_5$Ni$_2$In$_4$ below $T_C=30$~K~\cite{tyvanchuk2010magnetic}. Er$_5$Ni$_2$In$_4$ shows ferromagnetic ordering below
21~K~\cite{zhang2018investigation}. Neutron diffraction data reveal a complex magnetic structure in Er$_5$Ni$_2$In$_4$, containing both ferromagnetic and
antiferromagnetic components. The critical temperature of magnetic ordering is found to be 18.5~K and it is followed by
several additional magnetic transitions occurring at lower temperatures~\cite{GONDEK201210}. Tm$_5$Ni$_2$In$_4$ shows 
complex antiferromagnet character with a N\'{e}el temperature of 4.2~K~\cite{szytula2013magnetic} or 4.1~K \cite{SZYTULA2014149}.

In the RE$_{5}$Pd$_2$In$_4$ system, a ferromagnetic ordering is established in Tb$_{5}$Pd$_2$In$_4$ and Dy$_{5}$Pd$_2$In$_4$
below the Curie temperatures of 97~K and 88~K, respectively. As the temperature decreases, Tb$_{5}$Pd$_2$In$_4$ turns into an antiferromagnet,
while Dy$_{5}$Pd$_2$In$_4$ becomes a ferrimagnet. For Ho$_{5}$Pd$_2$In$_4$, a complex antiferromagnet emerges below its N\'eel temperature of
28.5~K, comprising two distinct antiferromagnetic components, one of which transitions into a ferrimagnetic state at lower temperatures.
A canted antiferromagnetic structure is found in Er$_{5}$Pd$_2$In$_4$ below $T_N=16.5$~K, with additional ferromagnetic component developing
at lower temperatures. Tm$_{5}$Pd$_2$In$_4$ shows antiferromagnetic behavior below $T_N=4.3$~K with two different antiferromagnetic components.
Analysis of the neutron diffraction data for RE$_{5}$Pd$_2$In$_4$ (RE = Gd–Er) indicates that the rare-earth moments at the 2a and 4g2 sites
order at higher temperatures than those at the 4g1 site~\cite{Baran2021}.
 
In the RE$_{5}$Pt$_2$In$_4$ system, DC and AC magnetic data reveal a transition from paramagnetic state to a ferromagnetic or
ferrimagnetic one for RE = Gd--Er. The Curie temperatures equal 76, 108, 23.5 and 12.6~K for RE = Gd, Tb, Ho and Er, respectively.
With decreasing temperature, additional anomalies in the magnetic susceptibility are observed for RE = Gd, Ho and Er. Tm$_{5}$Pt$_2$In$_4$
is found to be antiferromagnetic below $T_N = 4.1$~K~\cite{HAYYU2024175054}. The neutron diffraction data for RE = Tb--Er reveal
complex magnetic structures having both ferro- and antiferrromagnetic temperature-dependent components. As a result, the magnetic 
structures show a number of temperature-induced transformations for RE = Tb--Er. In contrary, a non-collinear antiferromagnetic
structure with no extra magnetic transitions is detected in Tm$_{5}$Pt$_2$In$_4$~\cite{Deptuch:lo5121}.


Up to now, the magnetocaloric properties of RE$_{5}$T$_2$In$_4$ (RE = rare earth element, T = transition metal element)
have been reported for selected compounds only, namely for RE$_{5}$Ni$_2$In$_4$ (RE = Dy, Ho, Er)~\cite{zhang2018investigation},
RE$_{5}$Pd$_2$In$_4$ (RE = Tb--Tm)~\cite{HAYYU2025417184} and RE$_{5}$Pt$_2$In$_4$
(RE = Gd--Tm)~\cite{HAYYU2024175054}.

Until now, there is only one paper on the physical properties of RE$_{5}$Rh$_2$In$_4$ (RE = rare earth element) limited
to basic crystallographic data solely. It motivated us to undertake the current study, in which we report detailed
crystallographic characterization and magnetic properties (including magnetocaloric effect) of the RE$_{5}$Rh$_2$In$_4$
(RE = Gd--Tm) intermetallics with multiple magnetic sublattices. The experimental results reported in this work are
compared with those reported previously for the isostructural compounds, including
RE$_5$Ni$_2$In$_4$~\cite{zaremba1991crystal,Ritter_2015,provino2012crystal,zhang2018investigation,tyvanchuk2010magnetic,GONDEK201210,szytula2013magnetic,SZYTULA2014149},
RE$_5$Pd$_2$In$_4$~\cite{Sojka200890,Baran2021, HAYYU2025417184} and
RE$_5$Pt$_2$In$_4$~\cite{zaremba2007,HAYYU2024175054, Deptuch:lo5121}.


\section{Materials and methods}

High-purity elements with at least 99.9 wt~\% of RE = Gd-Tm, as well as Rh and In have been selected as raw materials for arc-melting
in a titanium-gettered argon atmosphere with the RE:Rh:In atomic ratio of 5:2:4. The obtained ingots have been remelted for a few times
in order to ensure their homogeneity.

The phase composition of the samples has been identified by the X-ray powder diffraction at the room temperature 
using PANalytical X'Pert PRO diffractometer (Cu K$_{\alpha}$-radiation, Bragg-Brentano geometry,
angle interval of $2\theta = 10-100^\circ$, continuous scan mode, step size in $2\theta$ = 0.033$^\circ$, 300~s/step).
The FullProf program has been used for Rietveld analysis of the collected 
X-ray diffraction data set~\cite{Rodriguez-Carvajal_Physica_B_192,Rodriguez-Carvajal_Newsletter_26}.

For the magnetometric measurements, the powder samples have been encapsulated in plastic containers and glued with low-temperature
varnish in order to prevent random grain rotation during measurement.

The DC magnetic susceptibility has been measured with the Vibrating Sample Magnetometer (VSM) option of Physical Properties Measurement
System (PPMS) by Quantum Design, equipped with a superconducting magnet running up to 9~T (90~kOe). The data have been collected while
heating over the 1.85 - 390~K temperature interval, in presence of the magnetic field of 1~kOe. Zero Field Cooling (ZFC) regime
has been applied to the first measurement, while the Field Cooling (FC) one to the second run. Next, the sample has been heated above
its critical temperature of magnetic ordering, then demagnetized by the oscillating magnetic field, and after that cooled down to 1.9~K.
At the latter temperature an isothermal magnetization curve has been collected starting from zero field up to the maximum available
field of $\pm 90$~kOe. In order to study MCE, a number of magnetization vs. temperature curves have been measured under fix magnetic
field ranging from 10~kOe (1~T) up to 90~kOe (9~T) with a step of 10~kOe. Before each measurement, the sample has been demagnetized by
the oscillating magnetic field at temperature above respective critical temperature of magnetic ordering, and cooled down to 2~K
subsequently. Then the desired value of magnetic field has been set. The data have been collected from 2~K up to temperature well
above the transition to the paramagnetic state. The procedure has been repeated several times until reaching the maximum value
of magnetic field.

The AC magnetic susceptibility has been measured with the AC Measurement System (ACMS) option of PPMS. The data have been collected
from 1.9~K up to temperature above the respective critical temperature of magnetic ordering in the presence of oscillating magnetic
field with amplitude $H_{AC}=2$~Oe and direct magnetic field of $H_{DC}=1$~kOe. Based on these data, the real ($\chi$') and imaginary
($\chi$") parts of AC magnetic susceptibility have been determined for the frequencies of 500, 1000 and 2500~Hz.

\section{Crystal structure}

	\begin{figure*}
		\centering
		\includegraphics[scale=0.60, bb=0 0 851 567]{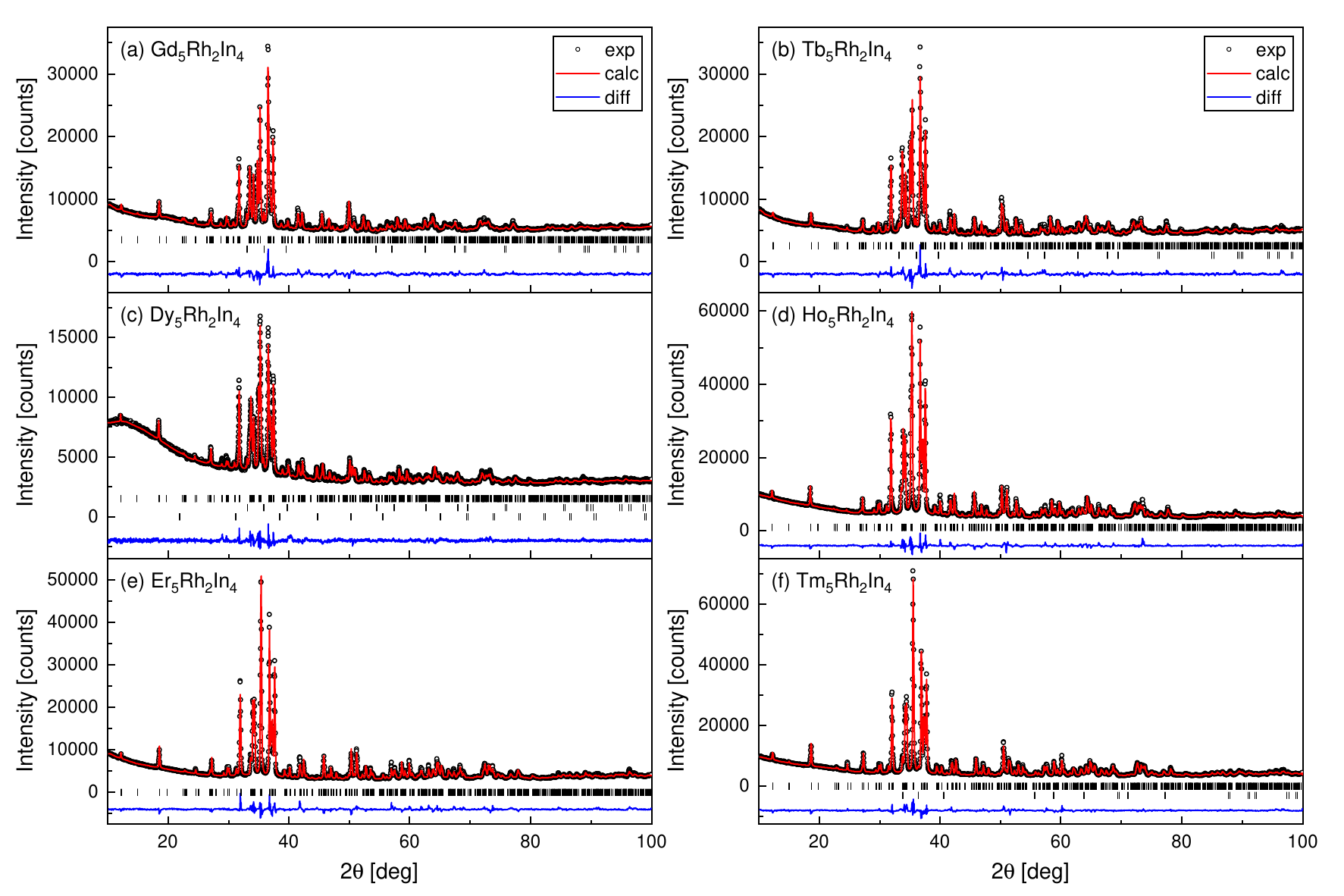}
		\caption{\label{fig:RE5Rh2In4_XRD_patterns}X-ray diffraction patterns at room temperature of RE$_{5}$Rh$_2$In$_4$: (a) RE = Gd, (b) RE = Tb,
			(c) RE = Dy (d) RE = Ho, (e) RE = Er, and (f) RE = Tm. The black solid circles show the
			experimental data, while the red lines show the Rietveld refinement results.
			The first row of vertical bars denotes Bragg reflection positions
			related to the main RE$_{5}$Rh$_2$In$_4$ phase, while the second
			one refers to RERh$_3$ (c) and elemental In (a, b, c, f). The blue line at the bottom shows the difference between the experimental data and the refinement results.}
	\end{figure*}

Fig.~\ref{fig:RE5Rh2In4_XRD_patterns} shows the result of Rietveld refinement of X-ray powder diffraction patterns of
\ch{RE5Rh2In4} (RE = Gd-Tm). The samples are found to contain at least 97~wt~\% of the main \ch{RE5Rh2In4} phase
with minor amounts of RERh$_3$ or elemental In for selected samples. The exact samples' compositions can be found
in Table~\ref{tbl:crystallographic_data}.

\begin{figure*}
		\centering
\includegraphics[width=0.75\textwidth]{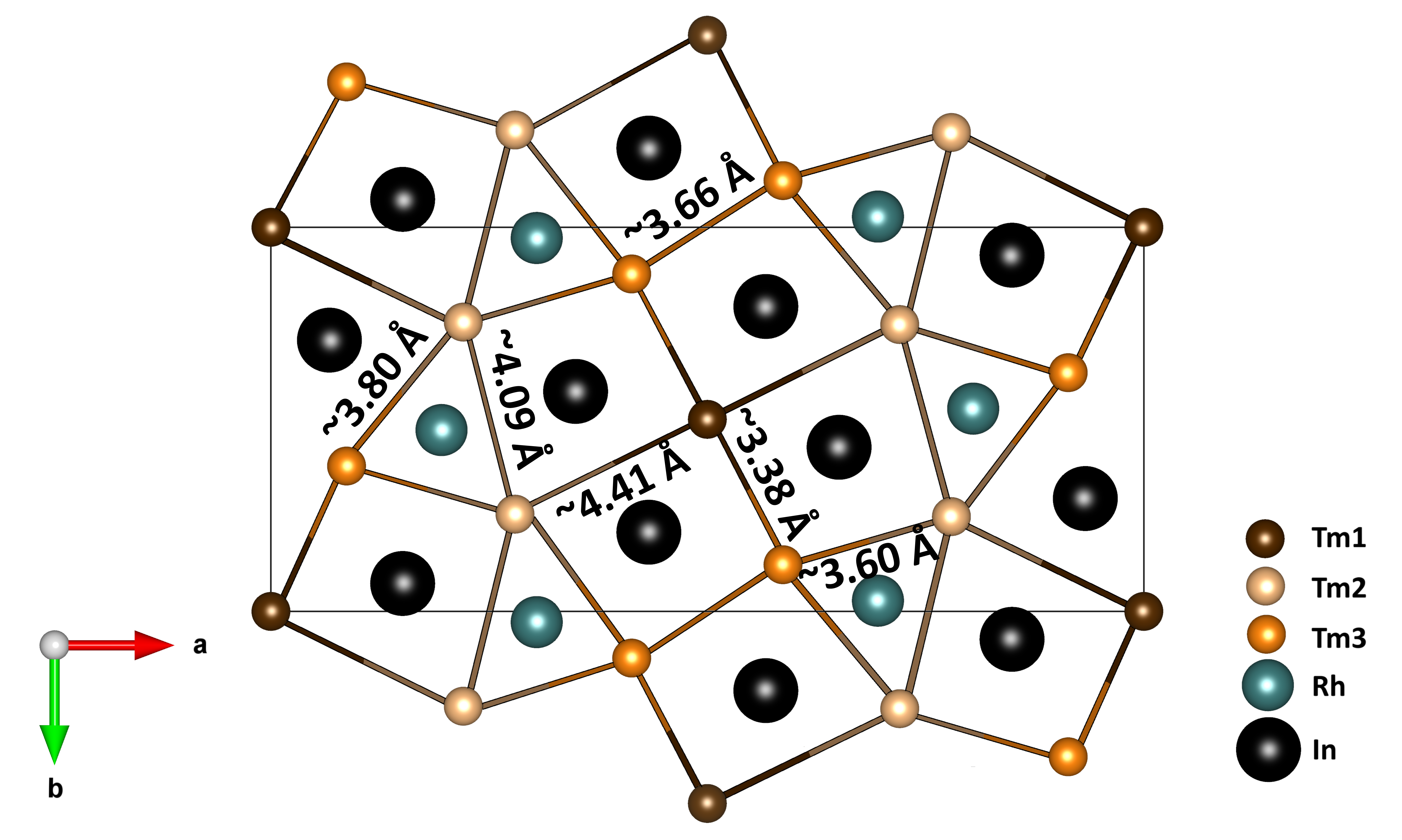}
\caption{\label{fig:RE5Rh2In4_crystal}The crystal unit cell of \ch{Tm5Rh2In4} projected onto the ab-plane.
	A network formed by the rare earth atoms and RE-RE interatomic distances are indicated.}
\end{figure*}

\ch{RE5Rh2In4} (RE = Gd-Tm) crystallize in the orthorhombic Lu$_{5}$Ni$_2$In$_4$-type of crystal structure (space group
\textit{Pbam}, No.~55), which has been reported for the first time by Zaremba et al.~\cite{zaremba1991crystal}. The structure
consists of the (001) planes ($z = 0$) containing rare earth atoms distributed among three non-equivalent Wyckoff sites, namely,
the 2a site (0,0,0) and two 4g sites $(x,y,0)$ with individual $x$ and $y$ atomic positional parameters. The rare earth
planes are separated by the planes consisting of two remaining elements $\left( z = \frac{1}{2} \right)$. 
The transition metal occupies one 4h site $\left( x,y,\frac{1}{2} \right)$, while the In atoms occupy two other 4h sites with individual $x$ and
$y$ atomic positional parameters. A projection of the crystal structure onto ab-plane, together with RE-RE interatomic
distances, is presented in Fig.~\ref{fig:RE5Rh2In4_crystal}. 

The values of lattice parameters, unit cell volumes and atomic positional parameters, as derived from Rietveld refinement of the
patterns shown in Fig.~\ref{fig:RE5Rh2In4_XRD_patterns}, are listed in Table ~\ref{tbl:crystallographic_data}. They agree well
with the ones reported previously~\cite{Zaremba2008}.



\begin{table*}[h!]
\begin{footnotesize}
\caption{\label{tbl:crystallographic_data}
Crystallographic data obtained from Rietveld refinement of the X-ray powder diffraction patterns at room temperature of RE$_{5}$Rh$_2$In$_4$ (RE = Gd-Tm) with Lu$_{5}$Ni$_2$In$_4$-type structure, space group $Pbam$, No. 55, where the RE atoms in crystal unit cell occupy three nonequivalent positions. The following factors $R_{profile}$, $R_{F}$, $R_{Bragg}$, and $\chi^{2}$ are reported. If present, the impurity phases are listed at the bottom of the table.}
\begin{tabular*}{0.98\textwidth}{@{\extracolsep{\fill}}lllllll}
\hline
RE & Gd & Tb & Dy & Ho & Er & Tm\\
\hline
$a$ [\r{A}] & 18.1754(36) & 18.1150(26) &  18.0596(23) &  18.0205(17) &  17.9783(17) &  17.9462(14)\\
$b$ [\r{A}] & 8.0062(16) & 7.9707(12) & 7.9529(10) & 7.9422(7) & 7.9265(7) & 7.8994(6)\\
$c$ [\r{A}] & 3.6627(9) & 3.6329(7) & 3.6078(6) & 3.5848(4) & 3.5649(5) & 3.5450(4)\\
$V$ [\r{A}$^3$] & 532.98(20) & 524.56(14) & 518.17(12) & 513.06(9) & 508.02(9) & 502.56(8)\\
RE1 at $2a$ (0, 0, 0) & 0 & 0 & 0 & 0 & 0 & 0\\
RE2 at $4g~(x, y, 0)$ & $x=0.2209(6)$ & 0.2216(6) & 0.2186(5) & 0.2205(4) & 0.2198(4) & 0.2205(3)\\ 
& $y=0.2494(17)$ & 0.2483(16) & 0.2422(14) & 0.2449(11) & 0.2464(11) & 0.2469(9)\\
RE3 at $4g~(x, y, 0)$ & $x=0.4157(7)$ & 0.4165(6) & 0.4153(2) & 0.4154(4) & 0.4140(4) & 0.4130(3)\\
& $y=0.1162(13)$ & 0.1217(13) & 0.1190(10) &  0.1206(9) & 0.1220(10) & 0.1205(7)\\
Rh at $4h~(x, y, \frac{1}{2})$ & $x=0.3063(6)$ & 0.3010(9) & 0.3011(8) & 0.3040(7) & 0.3037(5) & 0.3032(6)\\ 
& $y=0.0081(24)$ & 0.0148(22) & 0.0237(17) & 0.0272(13) & 0.0258(16) & 0.0275(13)\\
In1 at $4h~(x, y, \frac{1}{2})$ & $x=0.5714(8)$ & 0.5703(7) & 0.5693(6) & 0.5680(5) & 0.5682(6) & 0.5668(4)\\ 
& $y=0.2070(16)$ & 0.2063(14) & 0.2087(12) & 0.2076(9) & 0.2087(10) & 0.2057(8)\\
In2 at $4h~(x, y, \frac{1}{2})$ & $x=0.8500(8)$ & 0.8491(7) & 0.8493(6) & 0.8494(5) & 0.8491(5) & 0.8492(4)\\ 
& $y=0.0648(16)$ & 0.0717(16) & 0.0710(13) & 0.0700(10) & 0.0684(12) & 0.0723(10)\\
$R_{profile}$ [\%] & 1.99 & 2.14 & 1.71 & 2.53 & 2.89 & 2.47\\ 
$R_{F}$ [\%] & 7.43 & 6.23 & 6.69 & 4.34 & 6.42 & 4.29\\ 
$R_{Bragg}$ [\%] & 9.49 & 8.33 & 7.69 & 5.37 & 8.11 & 5.33\\ 
$\chi^{2}$ & 5.13 & 5.43 & 2.28 & 8.29 & 8.78 & 6.91\\
\hline
Sample composition &&&&&\\
RE$_{5}$Rh$_{2}$In$_{4}$ [wt \%] & 97.1(4) & 98.3(3) & 97.5(6) & 100 & 100 & 99.3(2)\\
RERh$_{3}$ [wt \%] &&& 0.4(5) &&&\\
In [wt \%] & 2.9(4) & 1.7(3) & 2.1(3) &&& 0.7(2)\\
\hline
\end{tabular*}
\end{footnotesize}
\end{table*}

\section{Results and Discussion}

\subsection{\label{Basic_magnetic_properties}Basic magnetic properties}

	\begin{figure*}
		\centering
		\includegraphics[scale=1.0, bb=0 0 493 596]{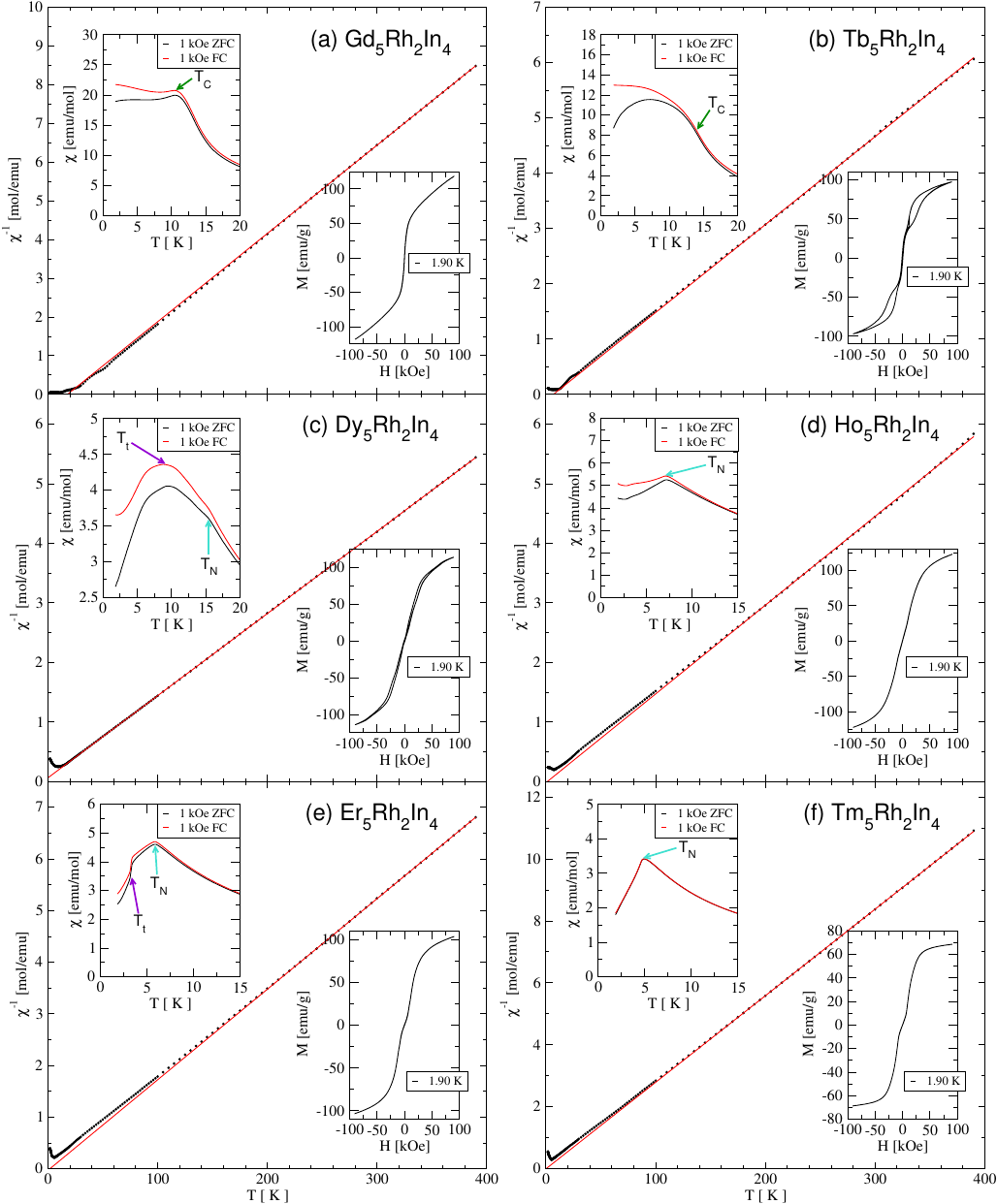}
		\caption{\label{fig:Reciprocal_X_vs_T}
DC reciprocal magnetic susceptibility with the fitted line representing the Curie-Weiss law (prime graph), the low-temperature ZFC and FC behavior measured at 1~kOe (upper insets), 
and the isothermal magnetization vs. external magnetic field loops at 1.9~K (lower insets) for RE$_{5}$Rh$_2$In$_4$: (a) RE = Gd, (b) RE = Tb, (c) RE = Dy (d) RE = Ho, (e) RE = Er,
and (f) RE = Tm. The transition temperatures listed in Table~\ref{tbl:DC_TC_TN_Tt_data} are indicated with arrows in the upper insets.}
	\end{figure*}

	\begin{figure*}
		\centering
		\includegraphics[scale=0.8, bb=0 0 588 596]{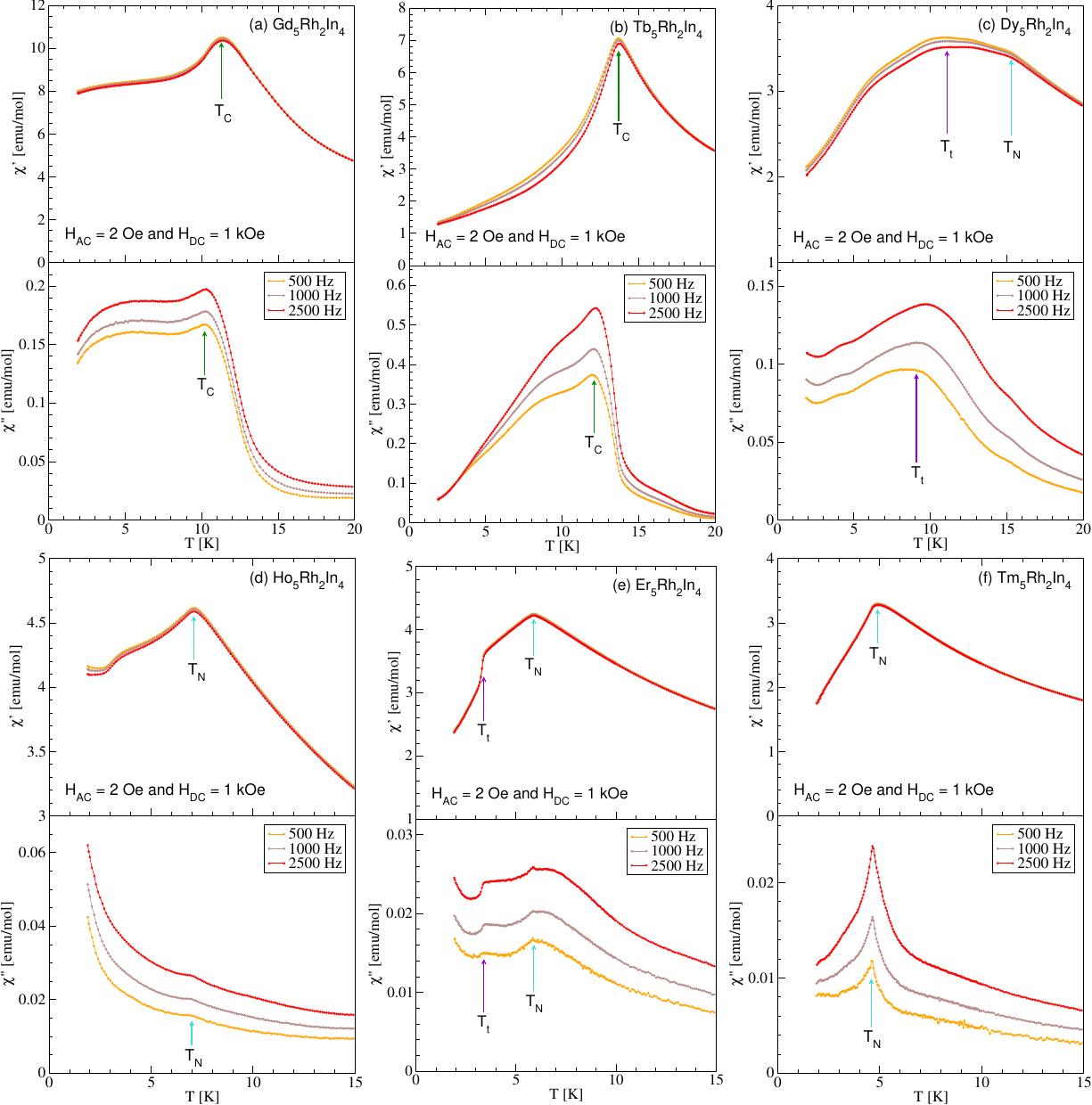}
		\caption{\label{fig:AC_magn_susc}
The real ($\chi$') and imaginary ($\chi$") parts of AC magnetic susceptibility taken at frequencies of 500~Hz, 1000~Hz and 2500~Hz for RE$_{5}$Rh$_2$In$_4$: (a) RE = Gd, (b) RE = Tb, (c) RE = Dy, (d) RE = Ho, (e) RE = Er, and (f) 
RE = Tm, respectively. The transition temperatures indicated with arrows are listed in Table~\ref{tbl:DC_TC_TN_Tt_data}.}
	\end{figure*}

	\begin{table*}
			\begin{flushleft}
				\normalsize
				\caption{\label{tbl:DC_TC_TN_Tt_data} Parameters characterizing magnetic order as derived from DC and AC magnetometric measurements:
				$T_{C}$ (Curie temperature), $T_{N}$ (N\'eel temperature), $T_{t}$ (temperatures of additional anomalies), $\theta_{p}$ (paramagnetic Curie temperature),
				$\mu_{eff}$ (effective magnetic moment), $\mu$ (magnetic moment in the ordered state), $H_{cr}$ (the critical field of metamagnetic transition)
				and $H_{coer}$ (the coercivity field) for RE$_{5}$Rh$_2$In$_4$ (RE = Gd--Tm). The $m$ index indicates a transition temperature corresponding to
				a maximum in the $\chi(T)$ curve, the $i$ index the one corresponding to an inflection point, while the $k$ index the one corresponding to a kink.
				The transition temperatures derived from the AC magnetometric measurements are based on the data at 1~kHz.
				The values of $\mu$, $H_{cr}$ and $H_{coer}$ have been determined from the data collected at 1.9~K in magnetic fields up to 90~kOe.
				}
			\end{flushleft}
			\vspace{-0.2 cm}
			\small
			\tabcolsep=0.04cm
			\begin{tabular*}{0.99\textwidth}{@{\extracolsep{3pt}}ccccccccccccccccc}
				\hline
RE & \multicolumn{3}{c}{$T_{C}$[K]} & \multicolumn{3}{c}{$T_{N}$[K]} & \multicolumn{3}{c}{$T_{t}$[K]} & $\theta_{p}[K]$ & \multicolumn{2}{c}{$\mu_{eff}[\mu_B]$} & \multicolumn{2}{c}{$\mu[\mu_B]$}
& $H_{cr}[kOe]$ & $H_{coer}[kOe]$ \\
\cline{2-4} \cline{5-7} \cline{8-10} \cline{12-13} \cline{14-15}
& \multicolumn{1}{l}{$\chi_{dc}$} & \multicolumn{1}{l}{$\chi'_{ac}$} & \multicolumn{1}{l}{$\chi''_{ac}$} & \multicolumn{1}{l}{$\chi_{dc}$} & \multicolumn{1}{l}{$\chi'_{ac}$} & \multicolumn{1}{l}{$\chi''_{ac}$}
& \multicolumn{1}{l}{$\chi_{dc}$} & \multicolumn{1}{l}{$\chi'_{ac}$} & \multicolumn{1}{l}{$\chi''_{ac}$} && Exp. & Theor. & Exp. & Theor. \\
\hline
Gd & 10.6$^{m}$ & 11.3$^{m}$ & 10.2$^{m}$ &&&&&&& +17.9 & 8.37 & 7.94 & 6.15 & 7.00 & 0.40 & 0.17\\
Tb & 14.2$^{i}$ & 13.7$^{m}$ & 12.1$^{m}$ &&&&&&& +6.9 & 10.02 & 9.72 & 5.08 & 9.00 & 28 & 0.76\\
Dy &&&& 15.4$^{k}$ & 15.3$^{k}$ && 9.5$^{m}$ & 11.1$^{m}$ & 9.1$^{m}$ & -4.5 & 10.77 & 10.65 & 6.03 & 10.00 & 9.3, 24 & 0.72\\
Ho &&&& 7.2$^{m}$ & 7.1$^{m}$ & 7.0$^{k}$ &&&& +1.4 & 10.36 & 10.61 & 6.54 & 10.00 &  10 & 0.14\\
Er &&&& 5.9$^{m}$ & 5.9$^{m}$ & 5.9$^{m}$ & 3.4$^{i}$ & 3.4$^{i}$ & 3.4$^{i}$ & +1.9 & 9.54 & 9.59 & 5.55 & 9.00 & 7.7, 12 & 0.22\\
Tm &&&& 4.9$^{m}$ & 4.9$^{m}$ & 4.6$^{m}$ &&&& +0.2 & 7.57 & 7.57 & 3.71 & 7.00 & 9.3 & \\
\end{tabular*}
	\end{table*}

Fig.~\ref{fig:Reciprocal_X_vs_T} presents temperature dependences of reciprocal magnetic susceptibilities collected at $H = 1$~kOe
for RE$_{5}$Rh$_2$In$_4$ (RE = Gd-Tm). Theory predicts that a material which contains solid magnetic moments and shows
spontaneous magnetic ordering at low temperatures should follow the Curie-Weiss law (which is an extension of the Curie’s law
with Weiss’s molecular field~\cite{kittel2004introduction,Mugiraneza2022}) at its paramagnetic temperature region, i.e. above
the critical temperature of magnetic ordering. The Curie-Weiss law is given by the following equation:

	\begin{equation}
		\chi= \frac{C}{T-\theta_p} \label{eq:curie-weiss} 
	\end{equation}

\noindent where $C$ is the Curie constant and $\theta_p$ is a parameter called Curie-Weiss temperature or paramagnetic Curie temperature.
The fitted Curie-Weiss formula at high temperature range for RE$_{5}$Rh$_2$In$_4$ (RE = Gd-Tm) is represented by solid lines in
Fig.~\ref{fig:Reciprocal_X_vs_T}. Based on the slopes and intercepts of the fitted lines, the effective magnetic moments ($\mu_{eff}$) and
paramagnetic Curie temperatures ($\theta_p$) have been determined. Their values are listed in Table~\ref{tbl:DC_TC_TN_Tt_data}.
It is worth noting that the effective magnetic moments for RE$_{5}$Rh$_2$In$_4$ (RE = Gd-Tm) are close to the corresponding theoretical values
predicted for the RE$^{3+}$ ions, indicating that the magnetism in the investigated compounds is solely or predominantly related to the
magnetic moments of the rare earth ions. The upper insets in Fig.~\ref{fig:Reciprocal_X_vs_T} show ZFC and FC temperature dependence
of DC magnetic susceptibility collected at 1~kOe, while the lower insets present isothermal magnetization loops taken at 1.9~K.
The magnetic moments at $T = 1.9$~K and $H = 90$~kOe reach ~88~\% of its theoretical RE$^{3+}$ ion value for RE = Gd and not more than
66~\% of corresponding free ion values for the remaining RE = Tb-Tm (see Table~\ref{tbl:DC_TC_TN_Tt_data}). Such a huge reduction of
magnetic moment in the ordered state is usually attributed to the interaction with the crystalline electric field (CEF). However, one
should also take into account that the lack of saturation at $T = 1.9$~K and $H = 90$~kOe also contributes to the reduction of the
experimental values.

Fig.~\ref{fig:AC_magn_susc} reveals the temperature behavior of the real and imaginary parts of the AC magnetic susceptibility collected
in presence of $H_{dc} = 1$~kOe in order to make the results consistent with those in Fig.~\ref{fig:Reciprocal_X_vs_T}.

The magnetic transition temperatures together with critical fields $H_{cr}$ related to metamagnetic transitions and
coercivity fields $H_{coer}$ are listed in Table~\ref{tbl:DC_TC_TN_Tt_data}. As the RE$_{5}$Rh$_2$In$_4$ (RE = Gd-Tm)
compounds show complex magnetic behavior, indicating the presence of both ferro- and antiferromagnetic components of
low-temperature magnetic structures, their basic magnetic properties are analyzed individually in the following
subsections.

\subsubsection{Gd$_{5}$Rh$_2$In$_4$}

With decreasing temperature, the DC magnetic susceptibility of Gd$_{5}$Rh$_2$In$_4$ has a slight maximum at about 11~K followed by a plateau
(see the upper inset in Fig.~\ref{fig:Reciprocal_X_vs_T}a). Usually, a maximum is characteristic of antiferromagnetic order. However,
there is a lot of evidence for the existence of the ferromagnetic component of the magnetic structure too, namely,
divergence of the ZFC and FC susceptibility curves below 11~K, positive value of the paramagnetic Curie temperature of +17.9~K,
shape of the isothermal magnetization loop collected at 1.9~K, which is typical of a soft ferromagnetic material with
coercivity field of 0.17~kOe at 1.9~K (see the lower inset in Fig.~\ref{fig:Reciprocal_X_vs_T}a) as well as distinct maximum
in the imaginary component of the AC magnetic susceptibility ($\chi''_{AC}$) (see Fig.~\ref{fig:AC_magn_susc}a). A weak antiferromagnetic
component of the magnetic structure is confirmed by metamagnetic transition occurring at the small critical field of 0.50~kOe
as derived from the primary isothermal magnetization curve taken at 1.9~K.
Summarizing, the results of magnetic measurements state that the low-temperature magnetic structure in Gd$_{5}$Rh$_2$In$_4$ 
consists of both ferro- and antiferromagnetic components with predominant ferromagnetic character.

\subsubsection{Tb$_{5}$Rh$_2$In$_4$}

DC magnetic susceptibility of Tb$_{5}$Rh$_2$In$_4$ is typical of a ferromagnet with inflection point at about 14~K, indicating
the Curie temperature and divergence of the ZFC and FC susceptibility curves below $T_C$. Predominant ferromagnetic character
of the magnetic order is also confirmed by a positive value of the paramagnetic Curie temperature of +6.9~K, coercivity field of
0.76~kOe at 1.9~K as well as distinct maximum in $\chi''_{AC}$ visible near $T_C$. An antiferromagnetic component of the
magnetic structure is also present as the primary isothermal magnetization curve (taken at 1.9~K) reveals a metamagnetic
transition at the critical field of 28~kOe. The shape of the isothermal magnetization loop at 1.9~K confirms coexistence
of ferro- and antiferromagnetic contributions to the magnetic order, as the magnetic field hysteresis at 1.9~K is relatively
small below the critical field of metamagnetic transition ($H_{cr}$) while it increases rapidly for the magnetic fields
exceeding $H_{cr}$, where the antiferromagnetic component of the magnetic structure is suppressed by application of
high magnetic field (see the lower inset in Fig.~\ref{fig:Reciprocal_X_vs_T}b).

\subsubsection{Dy$_{5}$Rh$_2$In$_4$}

With decreasing temperature, the DC magnetic susceptibility of Dy$_{5}$Rh$_2$In$_4$ shows two transitions: a small kink
at $T_N$ of about 15~K, followed by another magnetic transition manifesting itself by maximum at $T_t$ of about 10~K. 
Although a divergence of the ZFC and FC susceptibility curves, revealing a ferromagnetic contribution to the magnetis order,
is visible below $T_N$, there is only a very small anomaly in $\chi''_{AC}$ at $T_N$ indicating rather antiferromagnetic
character of magnetic order between $T_N$ and $T_t$. The ferromagnetic component of the magnetic structure is enhanced
at $T_t$, where $\chi''_{AC}$ reaches its local maximum. The isothermal magnetization curve collected at 1.9~K shows
features characteristic of both ferro- and antiferromagnetic ordering, namely, visible hysteresis with a coercivity 
field of 0.72~kOe (ferromagnetic character), while the primary magnetization curve reveals two subsequent metamagnetic
transitions with critical fields equal to 9.3 and 24~kOe, respectively (antiferromagnetic character). The predominant
nature of the magnetic interactions is antiferromagnetic, as the paramagnetic Curie temperature is negative
($\theta_{p} = -4.5$~K).

\subsubsection{Ho$_{5}$Rh$_2$In$_4$}

DC magnetic susceptibility of Ho$_{5}$Rh$_2$In$_4$ has a maximum typical of para- to antiferromagnetic transition
at about 7~K. The predominant antiferromagnetic character of magnetic order is confirmed by the shape of the
isothermal magnetization loop taken at 1.9~K with a distinct metamagnetic transition at $H_{cr} = 10$~kOe. A small
ferromagnetic component of the magnetic structure is also present, as the coercivity field of 0.14~kOe is observed at
1.9~K, a divergence of the ZFC and FC DC magnetic susceptibility curves is visible below the critical temperature
of the magnetic ordering, as well as there is a tiny anomaly in $\chi''_{AC}$ at $T_N$. The paramagnetic Curie
temperature is found to be +1.4~K. It is positive, but close to zero, indicating no clear dominance of either
ferro- or antiferromagnetic interactions.

\subsubsection{Er$_{5}$Rh$_2$In$_4$}

With decreasing temperature, the DC magnetic susceptibility of Er$_{5}$Rh$_2$In$_4$ shows a maximum at $T_N = 5.9$~K,
characteristic of para- to antiferromagnetic transition, followed by a rapid decrease at $T_t = 3.4$~K.
The isothermal magnetization loop at 1.9~K is typical of an antiferromagnet with two distinct metamagnetic
transitions at 7.7 and 12~kOe. Presence of small ferromagnetic contribution to the magnetic structure is
confirmed by the existence of coercivity field equal to 0.22~kOe at 1.9~K, divergence of the ZFC and FC
DC magnetic susceptibility curves below $T_N$ as well as tiny anomalies in $\chi''_{AC}$ present at both
$T_N$ and $T_t$. However, it should be emphasized that these anomalies reach very low values when
compared with the values of $\chi'_{AC}$ at the respective temperatures. The paramagnetic Curie temperature
of +1.9~K is positive, but too small to favor in a clear way predominant character of the magnetic order.

\subsubsection{Tm$_{5}$Rh$_2$In$_4$}

DC magnetic susceptibility of Tm$_{5}$Rh$_2$In$_4$ has a maximum typical of para- to antiferromagnetic transition
at about 5~K. The predominant antiferromagnetic nature of the magnetic order is confirmed by the shape of the 
isothermal magnetization loop at 1.9~K, characterized by a metamagnetic transition at 9.3~kOe and lack of
detectable coercivity field, as well as lack of visible divergence of the ZFC and FC DC magnetic susceptibility curves
below the critical temperature of the magnetic ordering. Nevertheless, a very small ferromagnetic component of the
magnetic structure may be present, as there is a tiny anomaly in $\chi''_{AC}$. However, it is more than
two orders of magnitude smaller than the corresponding anomaly in $\chi'_{AC}$.

\subsubsection{Discussion on the basic magnetic properties}

As shown above, the RE$_{5}$Rh$_2$In$_4$ (RE = Gd-Tm) intermetallics have complex magnetic properties characterized
by the presence of both ferro- and antiferromagnetic components of their low-temperature magnetic structures. Therefore,
the resultant magnetic structures arise from competition between magnetic interactions of different character.
Such a result is closely related to the complex crystal structure, where the rare earth atoms occupy three
non-equivalent Wyckoff sites and each site has its own arrangement of the neighbouring ions.

	\begin{figure}[ht!]
	    \centering
		\includegraphics[width=0.48\textwidth, bb=5 -6 570 405]{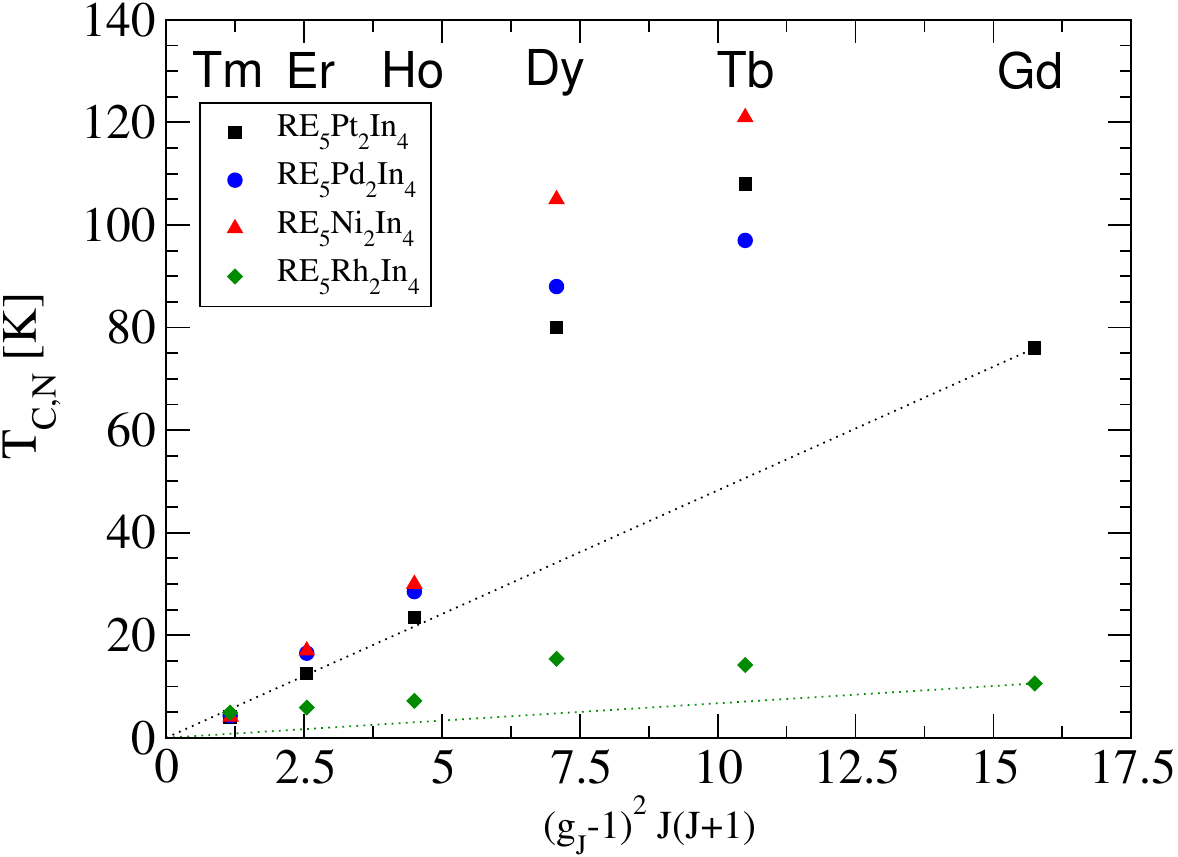}
		\caption{\label{fig:lande_factor_vs_Theta_P}Critical temperatures of magnetic ordering ($T_{C,N}$) vs. de Gennes factor for RE$_{5}$T$_2$In$_4$
		(RE = rare earth element, T = Ni~\cite{tyvanchuk2010magnetic,provino2012crystal,GONDEK201210,SZYTULA2014149,Ritter_2015},
		Pd~\cite{Baran2021}, Pt~\cite{HAYYU2024175054}, and Rh (this work)). Gd-based compounds are
		taken as a reference.}

	\end{figure}

According to the crystallographical data reported in Ref.~\cite{Zaremba2008}, the rare earth interatomic
distances in RE$_{5}$Rh$_2$In$_4$ (RE = Gd-Tm) exceed 3.35~\AA  (see the RE-RE interatomic distances
indicated in Fig.~\ref{fig:RE5Rh2In4_crystal}, as well as the values of the c lattice parameters listed in
Table~\ref{tbl:crystallographic_data}). Therefore, they are high enough to exclude any direct 
exchange interactions between the $4f$ subshells of the RE$^{3+}$ ions, which carry magnetic moments in the 
investigated compounds. In such a case, the indirect exchange interactions of the Ruderman-Kittel-Kasuya-Yosida
(RKKY)-type are expected in this family of compounds. One of the predictions of the RKKY theory is the so-called de Gennes 
scaling, which assumes direct proportionality between the critical temperature of magnetic ordering and the de Gennes
factor defined as $(g_J-1)^2J(J+1)$, where $g$ is a Land\'e factor, while $J$ is a total angular momentum of the
RE$^{3+}$ rare earth ion~\cite{De_Gennes_J_Phys_Radium}.

Fig.~\ref{fig:lande_factor_vs_Theta_P} presents the critical temperatures of magnetic ordering for RE$_{5}$T$_2$In$_4$
(T = Ni, Pd, Pt, Rh) plotted against the de Gennes factor. The dotted lines show theoretical de Gennes behavior
calculated for RE$_{5}$Pt$_2$In$_4$ and RE$_{5}$Rh$_2$In$_4$, where the critical temperatures of magnetic
ordering for the Gd-based compounds are available and taken as a reference. This is because, according to the
Hund's rules, the Gd$^{3+}$ ions do not interact with CEF. Distinct deviations from proportionality, visible for
all families of compounds, can be attributed to the influence of crystalline electric field (CEF)~\cite{NOAKES198235}.
Such a result is in agreement with values of the magnetic moments of RE$_{5}$Rh$_2$In$_4$ (RE = Gd-Tm)
in the magnetically ordered state (at $T = 1.9$~K and $H = 90$~kOe) as listed in Table~\ref{tbl:DC_TC_TN_Tt_data}.
The magnetic moment of Gd$_{5}$Rh$_2$In$_4$ reaches ~88~\% of the theoretical Gd$^{3+}$ ion value 
while the moments of the remaining compounds with RE = Tb-Tm are significantly reduced and do not exceed
66~\% of corresponding free ion values. Reduction of the magnetic moments for RE~=~Tb-Tm is another evidence
for interactions of the rare earth ions with CEF and emphasizes influence of this type of interactions
on the resultant magnetic structures.


For a selected rare earth element among Gd, Tb, Dy, Ho and Er, the critical temperature of the Rh-based RE$_{5}$T$_2$In$_4$ compound
is significantly lower than the temperatures of the Ni-, Pd- and Pt-based isostructural intermetallics
(see Fig.~\ref{fig:lande_factor_vs_Theta_P}). As differences in the lattice parameters for a fixed RE element are less than 1~\%
they can not explain such a large difference in the magnetic ordering temperatures. However, one should take into account
that Rh has one d-electron less in its unfilled d-subshell than Ni, Pd and Pt. Therefore, differences in the magnetic ordering
temperatures are rather related to the electronic structures of the compounds under investigation. Unfortunately, there are
no reports about electronic structure of the Rh-based RE$_{5}$T$_2$In$_4$ compounds, while there is one study reporting
the electronic structures of the isostructural Sc$_{5}$T$_2$In$_4$ (T = Ni, Pd and Pt)~\cite{Shafiq2020}. The latter
compounds have similar electronic structures. This result coincides with similar magnetic ordering temperatures
observed in RE$_{5}$T$_2$In$_4$ (RE = Tb--Tm; T = Ni, Pd, Pt) for a fixed RE element
-- see the data in Fig.~\ref{fig:lande_factor_vs_Theta_P}). Investigation of the correlation between magnetic ordering temperature
and electronic structure is possible in case of the Gd$_3$T compounds, which crystallize in the orthorhombic crystal structure
of the Fe$_3$C-type, where Gd occupies two non-equivalent Wyckoff sites (4c and 8d)~\cite{Kusz:ks0005}. Gd$_3$T are found
to order magnetically below 112 and 325~K for T = Rh and Pd, respectively~\cite{TALIK199587}. The valence band in Gd$_3$Rh
is formed by the hybridized Gd 5d and Rh 4d states (the latter ones located 2.1~eV below the Fermi
level)~\cite{TALIK1995795}, while the valence band in Gd$_3$Pd origins mainly from the Gd 5d states as the Pd 4d 
states have binding energy of 3.63~eV~\cite{TALIK1998183}.
Hybridization of the Gd 5d and Rh 4d states leads to partial depopulation of the Gd 5d states at the Fermi level.
Taking into account electronic structures in Gd$_3$Rh and Gd$_3$Pd, a much stronger depopulation effect is expected
for Gd$_3$Rh than for Gd$_3$Pd. As the RKKY theory of exchange interactions predicts that the critical temperature of
magnetic ordering is proportional to the density of states (DOS) at the Fermi level~\cite{De_Gennes_J_Phys_Radium},
the Rh-based rare earth intermetallics are expected to have lower magnetic ordering temperatures than the isostructural
Pd-based ones. Difference in location of the transition metal 4d states is also observed in the HoTX (T = transition metal;
X = p-electron element) intermetallics that crystallize in the orthorhombic crystal structure of the TiNiSi-type.
The UPS and XPS spectra reveal binding energy of the Rh 4d states equal to 3.0~eV in HoRhSi and to 2.4~eV in HoRhGe,
while the Pd 4d band in HoPdSn is 4.2~eV below the Fermi level~\cite{SZYTULA2003171}.


\subsection{Magnetocaloric effect}

	\paragraph{Magnetic entropy change and temperature averaged magnetic entropy change (TEC)}
\mbox{}
\bigskip

	\begin{figure*}
		\centering
		\includegraphics[scale=0.8, bb=0 0 517 596]{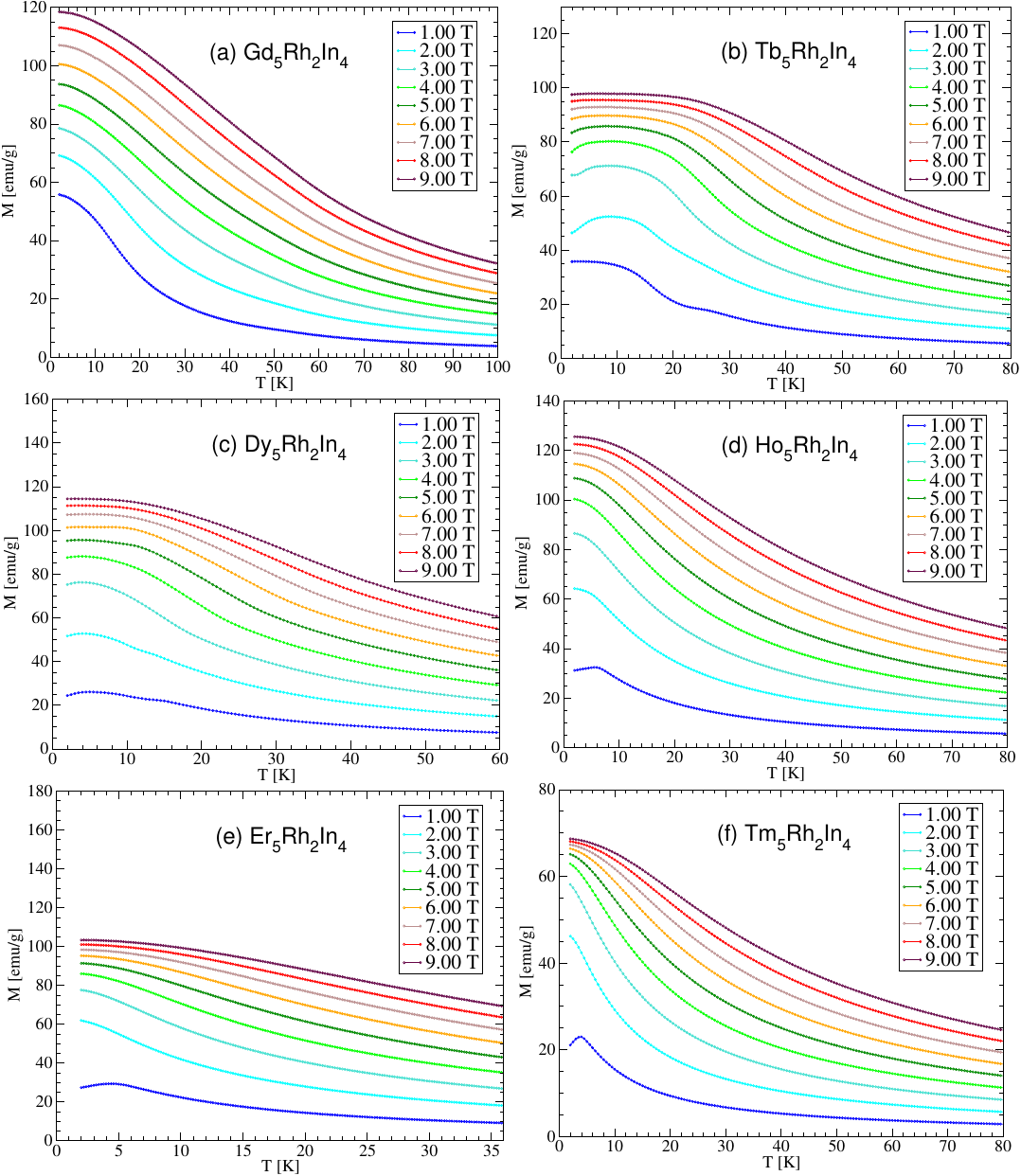}
		\caption{\label{fig:M_vs_T}ZFC magnetization vs. temperature curves in magnetic flux densities between 1~T
		and 9~T for RE$_{5}$Rh$_2$In$_4$: (a) RE = Gd, (b) RE = Tb, (c) RE = Dy, (d) RE = Ho (e) RE = Er, and (f) RE = Tm.}
	\end{figure*}

Fig.~\ref{fig:M_vs_T} shows temperature dependences of magnetization taken at selected magnetic fields. Based on these
data, magnetic entropy change can be derived by using the Maxwell relation as follows:

\begin{equation}
\Delta S_M(T,\Delta\mu_{0}H) = \int_{\mu_{0}H_{i}}^{\mu_{0}H_{f}} 
               \left( \frac{\partial M (\mu_{0}H,T)}{\partial T} \right)_{(\mu_{0}H)}  \,\mathrm{d}\mu_{0}H
\label{eq:dSm}
\end{equation}

\noindent where $\Delta\mu_{0}H$ represents the difference between the final $\mu_{0}H_{f}$ and initial $\mu_{0}H_{i}$ magnetic flux densities,
while $\left( \frac{\partial M (\mu_{0}H,T)}{\partial T} \right)_{(\mu_{0}H)}$ is the derivative of 
magnetization over temperature at a fixed magnetic flux density $\mu_{0}H$~\cite{tishin2003magnetocaloric}.
Usually, the initial magnetic flux density is fixed to zero. This convention is also used in the current study.

\begin{figure*}
	\centering
	\includegraphics[scale=0.8, bb=0 0 519 596]{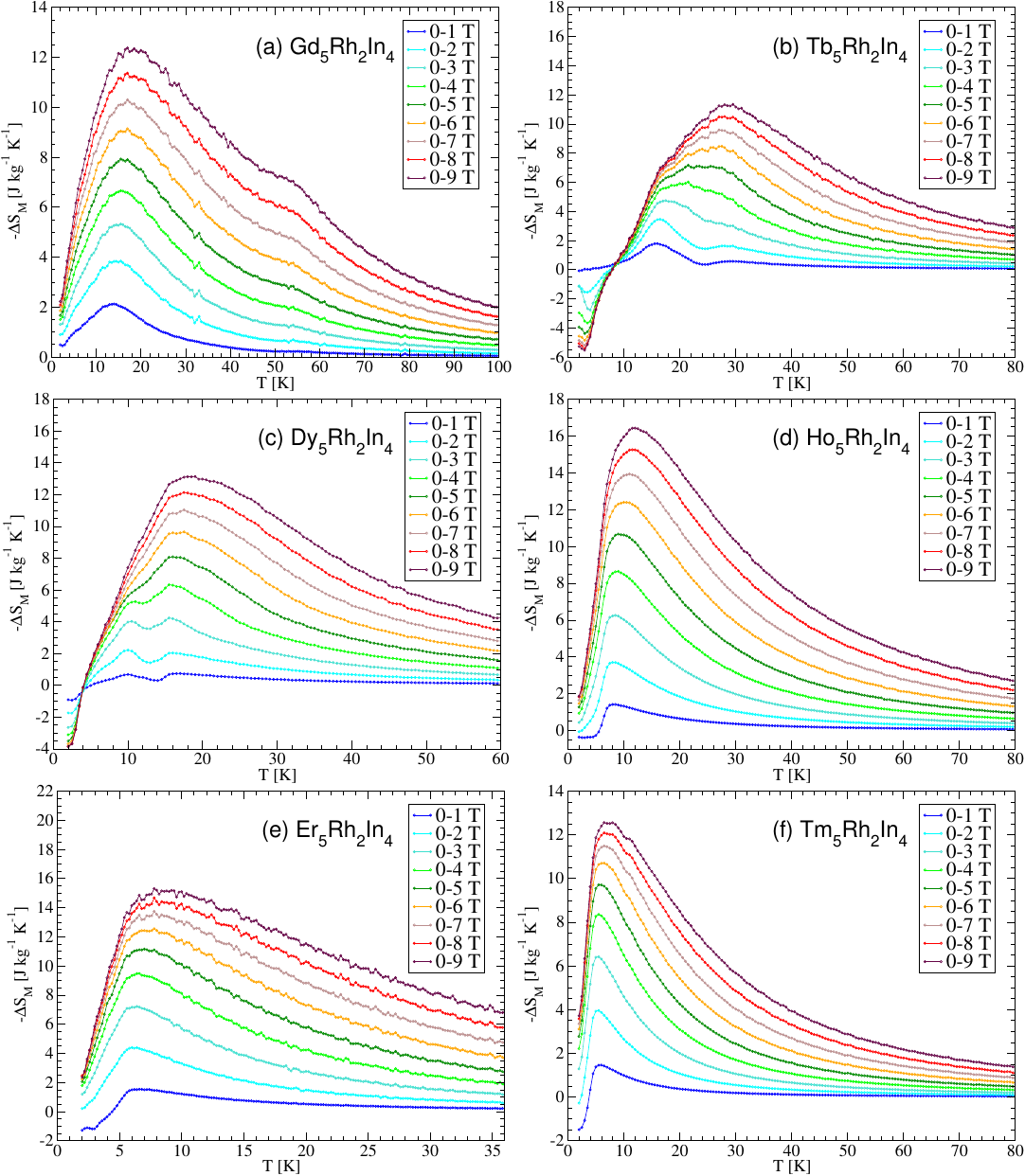}
	\caption{\label{fig:magn_entr_vs_T}Magnetic entropy change $-\Delta S_M^{max}$ vs. temperature $T$ as
	derived from $M(H,T)$ data shown in Fig.~\ref{fig:M_vs_T} under magnetic flux density changes $\Delta \mu_{0} H$ up to 
	0-9~T for RE$_{5}$Rh$_2$In$_4$: (a) RE = Gd, (b) RE = Tb, (c) RE = Dy, (d) RE = Ho (e) RE = Er, and (f) RE = Tm.}
\end{figure*}

Fig.~\ref{fig:magn_entr_vs_T} presents the calculated temperature dependencies of magnetic entropy change for RE$_{5}$Rh$_2$In$_4$
(RE = Gd-Tm) as derived from the data shown in Fig.~\ref{fig:M_vs_T} using Eq.~(\ref{eq:dSm}). The magnetic entropy change
reaches its maximum value in the vicinity of the respective temperature of transition from para- to magnetically ordered state.
Maximum magnetic entropy change equals 12.4, 11.3, 13.1, 16.4, 15.3, and 12.6~J$\cdot$kg$^{-1}\cdot$K$^{-1}$ under magnetic flux density
change of 0--9~T for RE = Gd, Tb, Dy, Ho, Er, and Tm, respectively. The temperatures at which the maximum value of $-\Delta S_M$
is found together with magnetic entropy changes, under selected values of magnetic flux density change, are listed in
Table~\ref{tbl:Tc_DeltaSM_RCP}. The same table also contains the values reported for other selected rare-earth-based compounds.
It is worth noting that magnetic entropy changes of RE$_{5}$Rh$_2$In$_4$ for RE = Tb and Dy are several times higher than those
observed in their Pd- and Pt-based analogues. This is not surprising as the critical temperatures of magnetic ordering
for RE$_{5}$Rh$_2$In$_4$ (RE = Tb, Dy) are remarkably smaller than those of RE$_{5}$T$_2$In$_4$ (RE = Tb, Dy; T = Pd, Pt).
Smaller critical temperature leads to significant magnetization change occuring in a narrower temperature range and
therefore maximizing  both $\left( \frac{\partial M (\mu_{0}H,T)}{\partial T} \right)_{(\mu_{0}H)}$ and $-\Delta S_M^{max}$
(see Eq.~(\ref{eq:dSm})). The same phenomenon explains the big differences in the $-\Delta S_M^{max}$ values for
Gd$_{5}$Rh$_2$In$_4$ and Gd$_{5}$Pt$_2$In$_4$ as well as for Dy$_{5}$Rh$_2$In$_4$ and Dy$_{5}$Ni$_2$In$_4$ (see
Table~\ref{tbl:Tc_DeltaSM_RCP}). The $-\Delta S_M^{max}$ values for \ch{RE5Rh2In4} (RE = Ho, Er and Tm) are found to be higher
than those of their Ni-, Pd- and Pt-based analogues, indicating stronger magnetization vs. temperature dependences in
the Rh-based compounds. The magnetic entropy changes found for \ch{RE5Rh2In4} (RE = Gd-Tm) are sizeable and comparable with
those reported for other rare-earth-based compounds with good magnetocaloric performance (see Table~\ref{tbl:Tc_DeltaSM_RCP}).
\ch{RE5Rh2In4} (RE = Gd-Tm) are therefore good candidates for application in low-temperature refrigeration techniques
based on adiabatic demagnetization.

An inverse MCE, which manifests itself in negative values of $-\Delta S_M^{max}$, is found at low temperatures and low values
of magnetic flux density change for selected \ch{RE5Rh2In4} (RE = Gd-Tm) intermetallics (see Fig.~\ref{fig:magn_entr_vs_T}).
Inverse MCE originates from the antiferromagnetic component of the magnetic structure, which is suppressed by increasing
temperature and/or increasing applied magnetic field. The inverse MCE is not observed for Gd$_{5}$Rh$_2$In$_4$. This result is
consistent with the magnetic data reported in section~\ref{Basic_magnetic_properties}, as Gd$_{5}$Rh$_2$In$_4$ has
predominantly ferromagnetic character.

The temperature-averaged magnetic entropy change (TEC) is a parameter reporting how magnetic entropy change
is maintained over a selected temperature span~\cite{griffith2018material-based}. TEC is calculated using the
following formula:

\begin{align} 
\MoveEqLeft TEC(\Delta T_\mathrm{lift},\Delta\mu_{0}H) =
\nonumber \\  &\frac{1}{\Delta T_\mathrm{lift}} \max_{T_\mathrm{mid}}
  \Bigg\{\int_{T_\mathrm{mid}-\frac{\Delta T_\mathrm{lift}}{2}}^{T_\mathrm{mid}+\frac{\Delta T_\mathrm{lift}}{2}} \Delta S_{\mathrm{M}}(T,\Delta\mu_{0}H)\,\mathrm{d}T\Bigg\}
\label{eq:tec}
\end{align}

\noindent where $T_\mathrm{mid}$ is the center temperature of the temperature span $\Delta T_\mathrm{lift}$.
The $T_\mathrm{mid}$ value is determined by finding the one that maximizes the integral in Eq.~\ref{eq:tec}.

\begin{figure*}
		\centering
		\includegraphics[scale=0.6, bb=0 0 713 595]{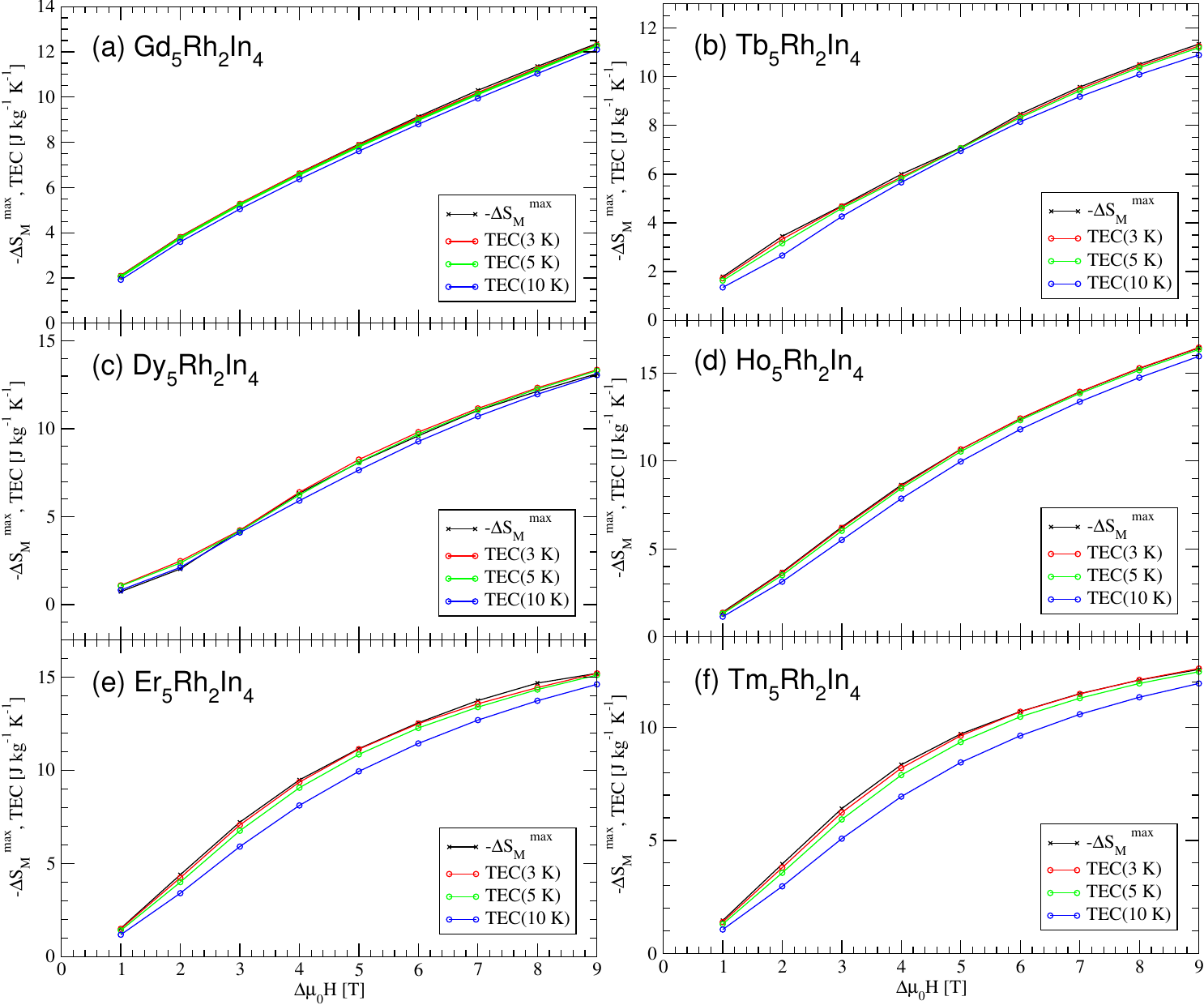}
		\caption{\label{fig:tec}The maximum magnetic entropy change $-\Delta S_M^{max}$ and temperature-averaged magnetic entropy change (TEC) of 3 K, 5 K, and 10 K plotted against magnetic flux density change $\Delta 
		\mu_{0}H$ for RE$_{5}$Rh$_2$In$_4$: (a) RE = Gd, (b) RE = Tb, (c) RE = Dy, (d) RE = Ho, (e) RE = Er, and (f) RE = Tm.}
	\end{figure*}

The TEC values calculated for 3, 5 and 10~K temperature spans for RE$_{5}$Rh$_2$In$_4$ (RE = Gd--Tm) are presented
in Fig.~\ref{fig:tec}. Within the 3~K span and magnetic flux density change of 0-9~T, TEC reaches 12.3~J$\cdot$kg$^{-1}\cdot$K$^{-1}$ (RE = Gd),
11.3~J$\cdot$kg$^{-1}\cdot$K$^{-1}$ (RE = Tb), 13.4~J$\cdot$kg$^{-1}\cdot$K$^{-1}$ (RE = Dy), 16.5~J$\cdot$kg$^{-1}\cdot$K$^{-1}$ (RE = Ho),
15.2~J$\cdot$kg$^{-1}\cdot$K$^{-1}$ (RE = Er) and 12.6~J$\cdot$kg$^{-1}\cdot$K$^{-1}$ (RE = Tm), respectively.
These values are quite close to the respective values of $-\Delta S_M^{max}$ (see Table~\ref{tbl:Tc_DeltaSM_RCP}).
TEC tends to decrease with increasing temperature span T$_{lift}$. This effect is especially noticeable
for the span of 10~K in the case of RE = Er and Tm. This is because \ch{RE5Rh2In4} (RE = Er, Tm) have the lowest
critical temperatures of magnetic ordering (see Table~\ref{tbl:DC_TC_TN_Tt_data}) and therefore the narrowest
maxima in the $-\Delta S_M(T)$ dependences (see Fig.~\ref{fig:magn_entr_vs_T}). However, it has to be mentioned that
even in case of RE = Er and Tm, the values of TEC(10~K) reach not less than 80~\% of the corresponding values
of $-\Delta S_M^{max}$ (see Fig.~\ref{fig:tec}), indicating that the magnetocaloric performance over the temperature
interval of 10~K is well maintened.

\bigskip

\bigskip

\paragraph{Relative Cooling Power (RCP) and Refrigerant Capacity (RC)}
\mbox{}
\bigskip

Besides $\Delta S_M^{max}$ and TEC, there are two more parameters used to describe magnetocaloric
performance of a given material. These are the relative cooling power (RCP)~\cite{gschneidner_pecharsky2000magnetocaloric_materials}
and refrigerant capacity (RC)~\cite{wood_potter1985general_analysis}. Both parameters correspond
to the amount of heat transfer between the cold and hot reservoirs in an ideal refrigeration cycle~\cite{Li_2016}. 
RCP is a product of maximum magnetic entropy change $-\Delta S_M^{max}$ and the full width 
at half maximum $\delta T_{FWHM}$ in the $-\Delta S_M(T)$ curve:

\begin{equation}
 RCP = -\Delta S_M^{max} \times \delta T_{FWHM}
\label{eq:rcp}
\end{equation}

While RC is numerically calculated by integrating the area under the {$-\Delta S_M(T)$} curve, taking the half maximum of the peak as the integration limits:

\begin{equation}
 RC = \int_{T_{1}}^{T_{2}} |\Delta S_M| \,\mathrm{d}T
\label{eq:rc}
\end{equation} 

\noindent where $T_1$ and $T_2$ are the lower and upper limits of the FWHM temperature range, respectively.

Under magnetic flux density change of 0-9~T, the RCP and RC values are found to be 656 and 498~J$\cdot$kg$^{-1}$ for RE = Gd,
442 and 340~J$\cdot$kg$^{-1}$ for RE = Tb, 363 and 358~J$\cdot$kg$^{-1}$ for RE = Dy, 522 and 407~J$\cdot$kg$^{-1}$ for RE = Ho,
437 and 335~J$\cdot$kg$^{-1}$ for RE = Er and 307 and 238~J$\cdot$kg$^{-1}$ for RE = Tm. The values corresponding to other
selected changes of magnetic flux density can be found in Table~\ref{tbl:Tc_DeltaSM_RCP}.

\bigskip

\paragraph{Summary of magnetocaloric performance of RE$_{5}$Rh$_2$In$_4$ (RE = Gd-Tm)}
\mbox{}
\bigskip

Table~\ref{tbl:Tc_DeltaSM_RCP} presents a number of parameters characterizing magnetocaloric performance of RE$_{5}$Rh$_2$In$_4$
(RE = Gd-Tm), including the temperature at which the maximum of magnetic entropy change is reached, together with the values of
$-\Delta S_{\mathrm{M}}^{\mathrm{max}}$, RCP and RC under magnetic flux density changes of 0-2, 0-5, 0-7 and 0-9~T.
For comparison, the parameters for the isostructural RE$_{5}$T$_2$In$_4$ (T = Ni, Pd, Pt) intermetallics, as well as selected 
well-performing rare-earth-based compounds are also included.

In terms of maximum magnetic entropy change ($-\Delta S_{\mathrm{M}}^{\mathrm{max}}$), RE$_{5}$Rh$_2$In$_4$ (RE = Gd-Tm) 
have the best magnetocaloric performance when compared to their isostructural RE$_{5}$T$_2$In$_4$ (T = Ni, Pd, Pt)
analogues. The magnetocaloric performance of RE$_{5}$Rh$_2$In$_4$ (RE = Gd-Tm) is comparable to the performance of
other well-performing rare-earth-based compounds, whose selected representatives are included in Table~\ref{tbl:Tc_DeltaSM_RCP}.

The temperatures at which maximum magnetic entropy change reaches its maximum vary for RE$_{5}$Rh$_2$In$_4$ (RE = Gd-Tm)
from 6.5~K (RE = Tm) up to 28~K (RE = Tb). Combining different RE$_{5}$Rh$_2$In$_4$ intermetallics makes it possible
to create a hybrid material showing good magnetocaloric performance from the liquid helium temperature up to about 30~K.
Moreover, combining selected RE$_{5}$Rh$_2$In$_4$ (RE = Gd-Tm) compounds with the ones well-performing at higher temperatures
(see selected representatives listed in Table~\ref{tbl:Tc_DeltaSM_RCP}) enables creating a material showing good
magnetocaloric performance from the liquid helium temperature up to the temperature of liquid nitrogen. Therefore,
RE$_{5}$Rh$_2$In$_4$ (RE = Gd-Tm) intermetallics, investigated in this work, are good candidates for application in
low-temperature refrigeration.

	\begin{table*}

\begin{footnotesize}
  \begin{flushleft}
  \normalsize
\caption{\label{tbl:Tc_DeltaSM_RCP}\ The temperature for maximum entropy changes, maximum entropy changes $-\Delta S_M^{max}$, RCP and RC values under magnetic flux density changes $\Delta \mu_0 H$ of 0--2~T, 0--5~T, 0--7~T, and 0--9~T for RE$_{5}$Rh$_2$In$_4$ (RE = Gd--Tm) and {selected rare-earth-based} compounds.} 
  \end{flushleft}
  \vspace{-0.2 cm}
  \footnotesize
  \begin{tabular*}{0.99\textwidth}{@{\extracolsep{\fill}}lllllllllllllll}
    \hline
    \hline
    Materials & Temp. for & \multicolumn{4}{c}{$-\Delta S_{\mathrm{M}}^{\mathrm{max}}$[J$\cdot$kg$^{-1}\cdot$K$^{-1}$]} & \multicolumn{4}{c}{RCP [J$\cdot$kg$^{-1}$]} & \multicolumn{4}{c}{RC [J$\cdot$kg$^{-1}$]} & Ref.\\
    & $-\Delta S_{\mathrm{M}}^{\mathrm{max}}$ [K] &&&&&&&&&&&&&\\
\multicolumn{1}{l}{} & \multicolumn{1}{l}{} & \multicolumn{1}{l}{0--2 T} & \multicolumn{1}{l}{0--5 T} & \multicolumn{1}{l}{0--7 T} & \multicolumn{1}{l}{0--9 T} &
	\multicolumn{1}{l}{0--2 T} & \multicolumn{1}{l}{0--5 T} & \multicolumn{1}{l}{0--7 T} & \multicolumn{1}{l}{0--9 T} & \multicolumn{1}{l}{0--2 T} & \multicolumn{1}{l}{0--5 T} & \multicolumn{1}{l}{0--7 T} & \multicolumn{1}{l}{0--9 T} & \\
    \hline
 Dy$_{5}$Ni$_2$In$_4$ & {103} & 1.8 & 3.6 & 4.7 & - & 49 & 178 & 286 & - & 37 & 130 & 209 & - & \citep{zhang2018investigation}\\
 Ho$_{5}$Ni$_2$In$_4$ & {19} & 2.6 & 7.1 & 10.1 & - & 84 & 298 & 458 & - & 66 & 234 & 352 & - & \citep{zhang2018investigation}\\
 Er$_{5}$Ni$_2$In$_4$ & 20 & 3.3 & 7.7 & 10.2 & - & 71 & 248 & 377 & - & 52 & 180 & 273 & - & \citep{zhang2018investigation}\\
 Gd$_{5}$Rh$_2$In$_4$ & 17 & 3.8 & 7.9 & 10.3 & 12.4 & 87 & 265 & 443 & 656 & 68 & 205 & 329 & 498 & this work\\
 Tb$_{5}$Rh$_2$In$_4$ & 28 & 3.5 & 7.2 & 9.6 & 11.3 & 34 & 199 & 315 & 442 & 26 & 161 & 246 & 340 & this work\\
 Dy$_{5}$Rh$_2$In$_4$ & 19 & 2.2 & 8.1 & 10.0 & 13.1 & 49 & 195 & 316 & 363 & 36 & 150 & 241 & 358 & this work\\
 Ho$_{5}$Rh$_2$In$_4$ & 12 & 3.7 & 10.7 & 13.9 & 16.4 & 53 & 227 & 368 & 522 & 40 & 176 & 287 & 407 & this work\\
 Er$_{5}$Rh$_2$In$_4$ & 8 & 4.4 & 11.1 & 13.7 & 15.3 & 49 & 191 & 311 & 437 & 36 & 147 & 236 & 335 & this work\\
 Tm$_{5}$Rh$_2$In$_4$ & 6.5 & 4.0 & 9.7 & 11.5 & 12.6 & 37 & 146 & 225 & 307 & 28 & 112 & 174 & 238 & this work\\ 
 Tb$_{5}$Pd$_2$In$_4$ & 62, 102 & 0.8 & 1.8 & 2.5 & 3.3 & 53 & 161 & 258 & 377 & 44 & 133 & 215 & 312 & \citep{HAYYU2025417184}\\
 Dy$_{5}$Pd$_2$In$_4$ & 22, 94 & 1.1 & 3.2 & 5.2 & 7.0 & 100 & 314 & 498 & 672 & 65 & 216 & 325 & 406 & \citep{HAYYU2025417184}\\
 Ho$_{5}$Pd$_2$In$_4$ & 22 & 2.5 & 7.2 & 10.0 & 12.6 & 94 & 326 & 489 & 661 & 81 & 269 & 403 & 541 & \citep{HAYYU2025417184}\\
 Er$_{5}$Pd$_2$In$_4$ & 17 & 3.3 & 7.7 & 10.1 & 12.1 & 103 & 287 & 403 & 528 & 91 & 227 & 314 & 409 & \citep{HAYYU2025417184}\\
 Tm$_{5}$Pd$_2$In$_4$ & 9 & 4.4 & 9.2 & 10.8 & 11.9 & 85 & 156 & 234 & 320 & 50 & 123 & 184 & 249 & \citep{HAYYU2025417184}\\
 Gd$_{5}$Pt$_2$In$_4$ & 78 & 1.0 & 2.2 & 3.0 & 3.7 & 58 & 172 & 261 & 359 & 48 & 139 & 209 & 290 & \citep{HAYYU2024175054}\\
 Tb$_{5}$Pt$_2$In$_4$ & 45, 110 & 0.8 & 1.7 & 2.5 & 3.4 & 82 & 198 & 303 & 428 & 57 & 165 & 248 & 340 & \citep{HAYYU2024175054}\\
 Dy$_{5}$Pt$_2$In$_4$ & 25, 95 & 1.1 & 2.9 & 4.7 & 6.3 & 45 & 290 & 250 & 363 & 33 & 201 & 180 & 263 & \citep{HAYYU2024175054}\\
 Ho$_{5}$Pt$_2$In$_4$ & 23 & 2.5 & 6.9 & 9.5 & 11.8 & 94 & 302 & 451 & 607 & 82 & 254 & 373 & 495 & \citep{HAYYU2024175054}\\
 Er$_{5}$Pt$_2$In$_4$ & 14 & 3.6 & 7.5 & 9.6 & 11.4 & 79 & 218 & 328 & 434 & 63 & 175 & 256 & 341 & \citep{HAYYU2024175054}\\
 Tm$_{5}$Pt$_2$In$_4$ & 8 & 3.2 & 7.7 & 9.2 & 10.2 & 34 & 125 & 189 & 260 & 27 & 98 & 150 & 205 & \citep{HAYYU2024175054}\\
 Gd$_{11}$Ni$_4$In$_9$ & 98 & 1.4 & 2.9 & 3.6 & - & 54 & 171 & 269 & - & 38 & 131 & 206 & - & \cite{ZHANG2021155863}\\
 Dy$_{11}$Ni$_4$In$_9$ & 19, 92 & 0.8 & 3.8 & 6.0 & - & 21 & 109 & 195 & - & 16 & 81 & 145 & - & \cite{ZHANG2021155863}\\
 Ho$_{11}$Ni$_4$In$_9$ & 16 & 1.8 & 8.8 & 12.4 & - & 32 & 198 & 353 & - & 24 & 152 & 269 & - & \cite{ZHANG2021155863}\\
 Gd$_{2}$In & 194 & 2.8 & - & - & - & - & - & - & - & - & - & - & - & \citep{BHATTACHARYYA20121239}\\ 
 Tb$_{2}$In & 165 & 3.5 & 6.6 & - & - & 200 & 660 & - & - & - & - & - & - &\citep{ZHANG2009396}\\
 Dy$_{2}$In & 130 & 4.6 & 9.2 & - & - & 230 & 736 & - & - & - & 545 & - & - &\citep{Zhang_2009}\\
 Dy$_{2}$In & 125 & - & - & 8.8 & - & - & - & - & - & - & - & - & - & \citep{YAO201937}\\
 Ho$_{2}$In & 85 & 5.0 & 11.2 & - & - & 125 & 560 & - & - & - & 360 & - & - & \citep{doi:10.1063/1.3130090, ZHANG2009396}\\
 Er$_{2}$In & 46 & 7.9 & 16.0 & - & - & - & - & - & - & - & 490 & - & - & \citep{ZHANG20112602}\\ 
 GdPt$_{2}$ & 28 & 1.7$^{a}$ & - & 6.2 & - & - & - & - & - & - & - & - & - & \citep{PhysRevB.74.132405}\\  
 DyPt$_{2}$ & 10.4 & 5 & 9.2 & 11.4$^{b}$ & - & - & - & - & - & 50 & 150 & - & - &  \citep{doi:10.1063/1.3253729}\\
 {GdRhIn} & $\sim$30 & {-} & {8} & 10.3 & - & - & - & - & - & 80 & {282} & - &  - & \citep{s24196326}\\
 GdRh & 24 & 13 & 22 & - & - & - & 534 & - & - & - & - & - & - & \citep{10.1063/1.4793775}\\
 Gd$_{3}$Rh & 112 & 5 & 9.2 & - & - & 143 & 473 & - & - & - & - & - & - & \citep{10.1063/1.3540664}\\
 Tb$_{3}$Rh & {94} & {11.6} & {17.9} & {21} & - & {226} & {594} & - & - & - & - & - &  - & \citep{TALIK2009L30}\\ 
 Ho$_{3}$Rh & {35} & - & {10} & - & - & - & - & - & - & - & {320} & - &  - & \citep{SHANG2020166055}\\  
 Gd$_{5}$Rh$_{4}$ & 16 & 5 & 14 & - & - & - & 374 & - & - & - & - & - & - & \citep{10.1063/1.4793775}\\
    \hline
    \hline
  \end{tabular*}
\end{footnotesize}
  \\
$^a$ $\Delta \mu_0 H$ = 0--1 T | $^b$ $\Delta \mu_0 H$ = 0--8 T 
\end{table*}

\section{Conclusions}

The RE$_{5}$Rh$_2$In$_4$ (RE = Gd-Tm) intermetallics crystallize in the orthorhombic Lu$_{5}$Ni$_2$In$_4$-type
crystal structure with the rare earth atoms occupying three non-equivalent Wyckoff sites. The compounds
order magnetically, with the critical temperatures of magnetic ordering ranging from 4.9~K (RE = Tm) up to
15.4~K (RE = Dy). The materials show complex magnetic properties indicating features characteristic of both
ferro- and antiferromagnetic orderings. The character of magnetic ordering changes with increasing number
of the 4f electrons from predominantly ferromagnetic in Gd$_{5}$Rh$_2$In$_4$ to predominantly
antiferromagnetic in Tm$_{5}$Rh$_2$In$_4$. The investigated compounds show at low temperatures quite good
magnetocaloric performance, comparable to that found in other well-performing rare-earth-based compounds.
Therefore, RE$_{5}$Rh$_2$In$_4$ (RE = Gd-Tm) are good candidates for application in low-temperature
refrigeration based on the adiabatic demagnetization technique.

\section*{Declaration of Competing Interest}

The authors declare no conflict of interest.

\section*{Acknowledgements}

The study was funded by ``Research support module'' as part of the ``Excellence Initiative -- Research University'' program
at the Jagiellonian University in Krak\'ow.

The research was carried out with the equipment purchased thanks to the financial support of the European Regional Development Fund
in the framework of the Polish Innovation Economy Operational Program (contract no.POIG.02.01.00-12-023/08).

	\bibliographystyle{elsarticle-num-names} 
	\bibliography{RE5Rh2In4_magn_arXiv_20250519_AH.bbl}

\end{document}